\documentclass[twocolumn,secnumarabic,amssymb, nobibnotes, aps, prd]{revtex4-2}
\usepackage[pdftex]{graphicx}
\usepackage{braket}

\begin{document}

\title{Search for ultralight dark matter from long-term frequency comparisons of optical and microwave atomic clocks}%

\author{Takumi Kobayashi$^{1}$}%
\email{takumi-kobayashi@aist.go.jp}
\author{Akifumi Takamizawa$^{1}$}
\author{Daisuke Akamatsu$^{1,2}$}
\author{Akio Kawasaki$^{1}$}
\author{Akiko Nishiyama$^{1}$}
\author{Kazumoto Hosaka$^{1}$}
\author{Yusuke Hisai$^{2\dag}$}
\author{Masato Wada$^{1}$}
\author{Hajime Inaba$^{1}$}
\author{Takehiko Tanabe$^{1}$}
\author{Masami Yasuda$^{1}$}
\affiliation{$^{1}$National Metrology Institute of Japan (NMIJ), National Institute of Advanced Industrial Science and Technology (AIST), 1-1-1 Umezono, Tsukuba, Ibaraki 305-8563, Japan \\ $^{2}$Department of Physics, Graduate School of Engineering Science, Yokohama National University, 79-5 Tokiwadai, Hodogaya-ku, Yokohama 240-8501, Japan\\$^{\dag}$Present address: Shimadzu Corporation, 3-9-4 Hikaridai, Seika-cho, Soraku-gun, Kyoto 619-0237, Japan}
\begin{abstract}
We search for ultralight scalar dark matter candidates that induce oscillations of the fine structure constant, the electron and quark masses, and the quantum chromodynamics energy scale with frequency comparison data between an $^{171}$Yb optical lattice clock and a $^{133}$Cs fountain microwave clock that span 298 days with an uptime of 15.4 $\%$. New limits on the couplings of the scalar dark matter to electrons and gluons in the mass range from $10^{-22}$ eV/$c^{2}$ to $10^{-20}$ eV/$c^{2}$ are set, assuming that each of these couplings is the dominant source of the modulation in the frequency ratio. The absolute frequency of the $^{171}$Yb clock transition is also determined as $518\,295\,836\,590\,863.69(28)$ Hz, which is one of the important contributions towards a redefinition of the SI second.
\end{abstract}
\maketitle

While the existence of dark matter is indicated by various astrophysical observations \cite{Bertone2018}, its constituents have not been conclusively detected in laboratory experiments. In such experiments, dark matter candidates with particle physics motivation are studied particularly well. Weakly interacting massive particles in the mass range from $1$ GeV/$c^{2}$ to $10^{3}$ GeV/$c^{2}$ ($c$: speed of light) have been searched for in various experiments \cite{Aprile2018,Akerib2017,Meng2021,Aaboud2018,Sirunyan2018,Aprile2019}. Quantum chromodynamics (QCD) axions and axion-like pseudoscalar particles and fields with masses of $\lesssim10$ eV/$c^{2}$ have also attracted considerable attention. Limits have been set on the interactions between the axions and the standard model particles such as electrons, photons, gluons, nucleons, and antiprotons \cite{Abel2017,Cheng2021,Roussy2021,Smorra2019,Bartram2021,Garcon2019}. The range of the search broadens as different kinds of technologies are applied to dark matter searches. 

Other possible candidates for dark matter are dilaton-like ultralight scalar fields with their masses $m_{\varphi}$ far below 1 eV/$c^{2}$, down to $10^{-22}$ eV/$c^{2}$ required by de Broglie wavelength of the scalar field being smaller than the sizes of dwarf galaxies \cite{Hu2000}. Such an ultralight bosonic field as dark matter has large occupation numbers per mode and behaves as a classical wave with a frequency proportional to $m_{\varphi}$ \cite{Safronova2018}. If the scalar field couples to the standard model fermions and gauge bosons \cite{Damour2010}, it induces coherent oscillations of fundamental constants such as the fine structure constant $\alpha$ and the electron mass $m_{e}$ \cite{Arvanitaki2015}. The scalar field can also form macroscopic clumps called topological defects \cite{Vilenkin1985}, leading to transient variations of fundamental constants by the passages of such clumps \cite{Derevianko2014,Wcislo2016,Roberts2017,Wcislo2018,Roberts2020}.

The variation of fundamental constants \cite{Ashby2007,Blatt2008,Rosenband2008,Guena2012,Leefer2013,Godun2014,Huntemann2014,Peil2013,McGrew2019,Schwarz2020,Lange2021} and thus the existence of the ultralight scalar fields can be detected with frequency ratios of atomic clocks based on different atomic species and transitions. The periodic oscillations of the frequency ratios have so far been searched for by clock comparisons involving optical and microwave clocks with different sensitivities to fundamental constants. Previous comparisons between two optical clocks took advantage of the state-of-the-art short-term frequency stabilities \cite{Oelker2019} and accuracies \cite{Bloom2014,McGrew2018,Brewer2019} to search for $\alpha$ oscillations in the frequency range of $\lesssim0.1$ Hz and set stringent bounds on coupling constants of the scalar fields to photons in the mass range of $m_{\varphi}\lesssim10^{-16}$ eV/$c^2$ \cite{Kennedy2020,Beloy2021}. Frequency ratios of two microwave clocks are sensitive to the couplings to quarks and gluons as well as photons. Although microwave clocks have relatively low short-term stabilities, they are operated for long periods with high uptimes and contribute to searches in low frequency (i.e., mass) regions. A comparison between Rb and Cs fountain microwave clocks for 6 years yielded stringent limits on the couplings to quarks and gluons in the mass range of $m_{\varphi}\lesssim10^{-21}$ eV/$c^2$ \cite{Hees2016}.

The unique feature of comparisons between optical and microwave clocks is the capability of the observation of the oscillation of $m_{e}$ induced by the scalar field couplings to electrons \cite{Arvanitaki2015}. The longest search for oscillations in the optical to microwave ratio was performed in a comparison between a Si optical cavity and a hydrogen maser (H maser) over 33 days \cite{Kennedy2020}. Assuming that the scalar fields predominantly couple to electrons, this report put constraints on the coupling between the scalar field and $m_{e}$ in the mass region from $m_{\varphi}=10^{-21}$ eV/$c^{2}$ to $10^{-18}$ eV/$c^{2}$. To extend the search to the lowest mass limit of $10^{-22}$ eV/$c^{2}$, longer measurement periods are required. It is also desirable to compare an optical lattice clock or a single ion optical clock with a fountain microwave clock to improve the dark matter detection sensitivity that is limited by the frequency instability of the Si/H measurement induced by flicker and random-walk frequency noises at long averaging times \cite{Milner2019}.  

In this paper, we report on a search for the oscillating scalar dark matter fields with frequency comparison data between an $^{171}$Yb optical lattice clock and a $^{133}$Cs fountain clock that span 298 days with an uptime of 15.4 $\%$.
The main technical advantage in our search is the robustness of the Yb optical lattice clock which can be operated with high uptimes for several months \cite{Kobayashi2020}. We establish improved constraints on the couplings of the scalar field to electrons and gluons in the mass range from $m_{\varphi}=10^{-22}$ eV/$c^{2}$ to $10^{-20}$ eV/$c^{2}$, assuming that each of these couplings is the dominant source of the oscillation. In addition to the dark matter search, we also provide the absolute frequency of the Yb clock transition, which is one of the important contributions towards a redefinition of the SI second \cite{Hong2016,Riehle2018}.

We consider linear couplings between the scalar field and the standard model particles described by the interaction Lagrangian \cite{Damour2010,Hees2016,Arvanitaki2015}
\begin{eqnarray}
\mathcal{L_{\mathrm{Int}}} &=&\varphi \Big[\frac{d_{e}}{4\mu_{0}}F_{\mu\nu}F^{\mu\nu} - \frac{d_{g}\beta_{3}}{2g_{3}}F^{A}_{\mu\nu}F^{A\mu\nu}\nonumber\\
&&-c^{2}\sum_{i=e,u,d}(d_{m_{i}}+\gamma_{m_{i}}d_{g})m_{i}\bar{\psi_{i}}\psi_{i} \Big],
\label{lagrangianeq}
\end{eqnarray}
where $\varphi$ denotes a dimensionless scalar field relative to the Planck scale \cite{Damour2010,Hees2016}, $F_{\mu\nu}$ the electromagnetic tensor, $\mu_{0}$ the magnetic permeability, $F^{A}_{\mu\nu}$ the gluon field strength tensor, $g_{3}$ the QCD coupling constant, $\beta_{3}$ the $\beta$ function for the running of $g_{3}$, $\gamma_{m_{i}}$ the anomalous dimension describing the running of the mass $m_{i}$ of the QCD-coupled fermion, and $\psi_{i}$ the fermion spinor. Five dimensionless coefficients $d_{e}$, $d_{m_{e}}$, $d_{m_{u}}$, $d_{m_{d}}$, and $d_{g}$ are coupling coefficients of the scalar field to photons, electrons, up and down quarks, and gluons, respectively. These couplings result in the following linear dependence of the fundamental constants with respect to the scalar field \cite{Damour2010}
\begin{eqnarray}
\alpha(\varphi)&=&\alpha(1+d_{e}\varphi),\quad m_{e}(\varphi)=m_{e}(1+d_{m_{e}}\varphi), \nonumber\\
m_{q}(\varphi)&=&m_{q}(1+d_{m_{q}}\varphi), \quad \Lambda_{\mathrm{QCD}}(\varphi)=\Lambda_{\mathrm{QCD}}(1+d_{g}\varphi),\nonumber\\
\label{xphi}
\end{eqnarray}
where $\Lambda_{\mathrm{QCD}}$ denotes the QCD energy scale, $m_{q}=(m_{u}+m_{d})/2$, and $d_{m_{q}}=(d_{m_{u}}m_{u}+d_{m_{d}}m_{d})/(m_{u}+m_{d})$.

With the variations of the fundamental constants, the fractional frequency ratio of two atomic clocks A and B changes such that 
\begin{equation}
\frac{\delta (f_{\mathrm{A}}/f_{\mathrm{B}})}{f_{\mathrm{A}}/f_{\mathrm{B}}} = k_{\alpha}\frac{\delta\alpha}{\alpha}+k_{m_{e}}\frac{\delta(m_{e}/\Lambda_{\mathrm{QCD}})}{m_{e}/\Lambda_{\mathrm{QCD}}}+k_{m_{q}}\frac{\delta(m_{q}/\Lambda_{\mathrm{QCD}})}{m_{q}/\Lambda_{\mathrm{QCD}}},
\label{ratio}
\end{equation}
where $k_{\alpha}$, $k_{m_{e}}$, and $k_{m_{q}}$ are sensitivity coefficients \cite{Flambaum2004}. For the Yb/Cs ratio, atomic and nuclear structure calculations yield $k_{\alpha}=-2.52$, $k_{m_{e}}=-1$, and $k_{m_{q}}=0.046$ \cite{Flambaum2006,Flambaum2004PRA,Dinh2009}. Note that $m_{e}/\Lambda_{\mathrm{QCD}}$ is used in Eq.~(\ref{ratio}) instead of the proton-to-electron mass ratio in conventional formula used in previous atomic clock experiments \cite{Ashby2007,Blatt2008,Rosenband2008,Guena2012,Leefer2013,Godun2014,Huntemann2014,Peil2013,McGrew2019,Schwarz2020,Lange2021}. From Eqs.~(\ref{xphi}) and (\ref{ratio}), the frequency ratio fluctuation caused by the scalar field is given by 
\begin{equation}
\frac{\delta (f_{\mathrm{Yb}}/f_{\mathrm{Cs}})}{f_{\mathrm{Yb}}/f_{\mathrm{Cs}}} = [-2.52d_{e}-d_{m_{e}}+0.046d_{m_{q}}+0.954d_{g}]\varphi.
\end{equation}
When the scalar field oscillates as $\varphi(t)\sim\varphi_{0}\cos(2\pi ft)$ with a frequency given by the Compton frequency $f=m_{\varphi}c^{2}/h$ ($h$: Planck constant), and carrys an energy density of $\rho_{\varphi}=c^{2}\pi f^{2}\varphi_{0}^{2}/(2G)$ ($G$: Newtonian constant of gravitation) \cite{Hees2016}, the frequency ratio oscillates such that $\delta (f_{\mathrm{Yb}}/f_{\mathrm{Cs}})/(f_{\mathrm{Yb}}/f_{\mathrm{Cs}})\sim A\cos(2\pi ft)$. Assuming that $\rho_{\varphi}$ consists of the local dark matter density $\rho_{\mathrm{DM}}\sim0.3$ GeV/cm$^{3}$, the relationship between the amplitude $A$ and the coupling coefficients is obtained as
\begin{equation}
A=[-2.52d_{e}-d_{m_{e}}+0.046d_{m_{q}}+0.954d_{g}]\sqrt{\frac{2G\rho_{\mathrm{DM}}}{\pi f^{2}c^{2}}}.
\label{darkmatteramp}
\end{equation}
The goal of the analysis is to estimate the amount of $A$ in the experimental data. 

\begin{figure}[t]
\includegraphics[scale=0.4,angle=0]{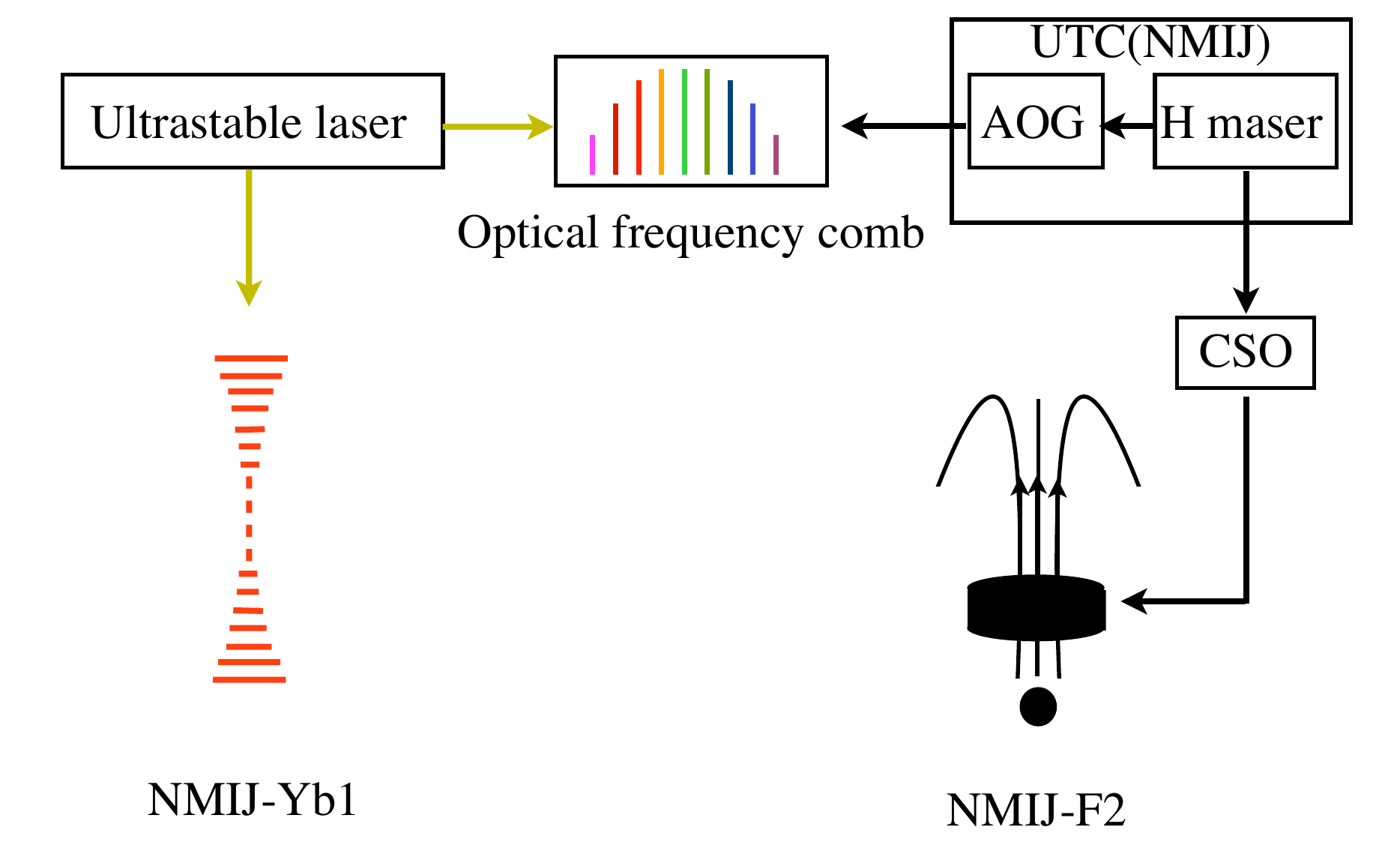}
\caption{Experimental setup. AOG:  auxiliary output generator, CSO: cryogenic sapphire oscillator.}
\label{experimentalsetup}
\end{figure}

The experimental setup is schematically described in Fig.~\ref{experimentalsetup}. At National Metrology Institute of Japan (NMIJ), we developed two atomic clocks: an Yb optical lattice clock (NMIJ-Yb1)  and a Cs fountain clock (NMIJ-F2).  The details of the experimental apparatuses of NMIJ-Yb1 and NMIJ-F2 are described in Refs.~\cite{Kobayashi2018,Kobayashi2020,Takamizawa2015,Takamizawa2021}, and thus we here provide a brief description. NMIJ-Yb1 utilizes the $^{1}$S$_{0}-^{3}$P$_{0}$ electronic transition of $^{171}$Yb at 518 THz, and NMIJ-F2 the $\ket{F=3,m_{F}=0}\to\ket{F=4,m_{F}=0}$ hyperfine transition of the ground state of $^{133}$Cs at 9.2 GHz. These two clocks are compared through a H maser which generates the Coordinated Universal Time of NMIJ (UTC(NMIJ)) with an auxiliary output generator (AOG). The frequency of NMIJ-Yb1 is compared with that of UTC(NMIJ) by counting a beat frequency between an ultrastable laser for probing the transition of Yb and an optical frequency comb \cite{Inaba2006} which is phase locked to UTC(NMIJ). NMIJ-Yb1 is then linked to the H maser with the AOG frequency measured by a time-interval counter. The frequency of NMIJ-F2 is compared with that of the H maser via an ultrastable cryogenic sapphire oscillator (CSO) \cite{Takamizawa2014} used as a local oscillator of NMIJ-F2.
Both NMIJ-Yb1 and NMIJ-F2 are operated nearly continuously. The operation of NMIJ-Yb1 is supported by a reliable laser system based on a frequency comb \cite{Hisai2019}, automatic laser relock schemes \cite{Kobayashi2019}, and remote controlling systems. The systematic frequency shifts of NMIJ-Yb1 and NMIJ-F2 are compensated according to methods described in Refs.~\cite{Kobayashi2018,Kobayashi2020,Takamizawa2015,Takamizawa2021}. Some of the systematic shifts of NMIJ-Yb1 are reevaluated as described in the Supplemental Material \cite{supplemental}. 

\begin{figure}[t]
\includegraphics[scale=0.43,angle=0]{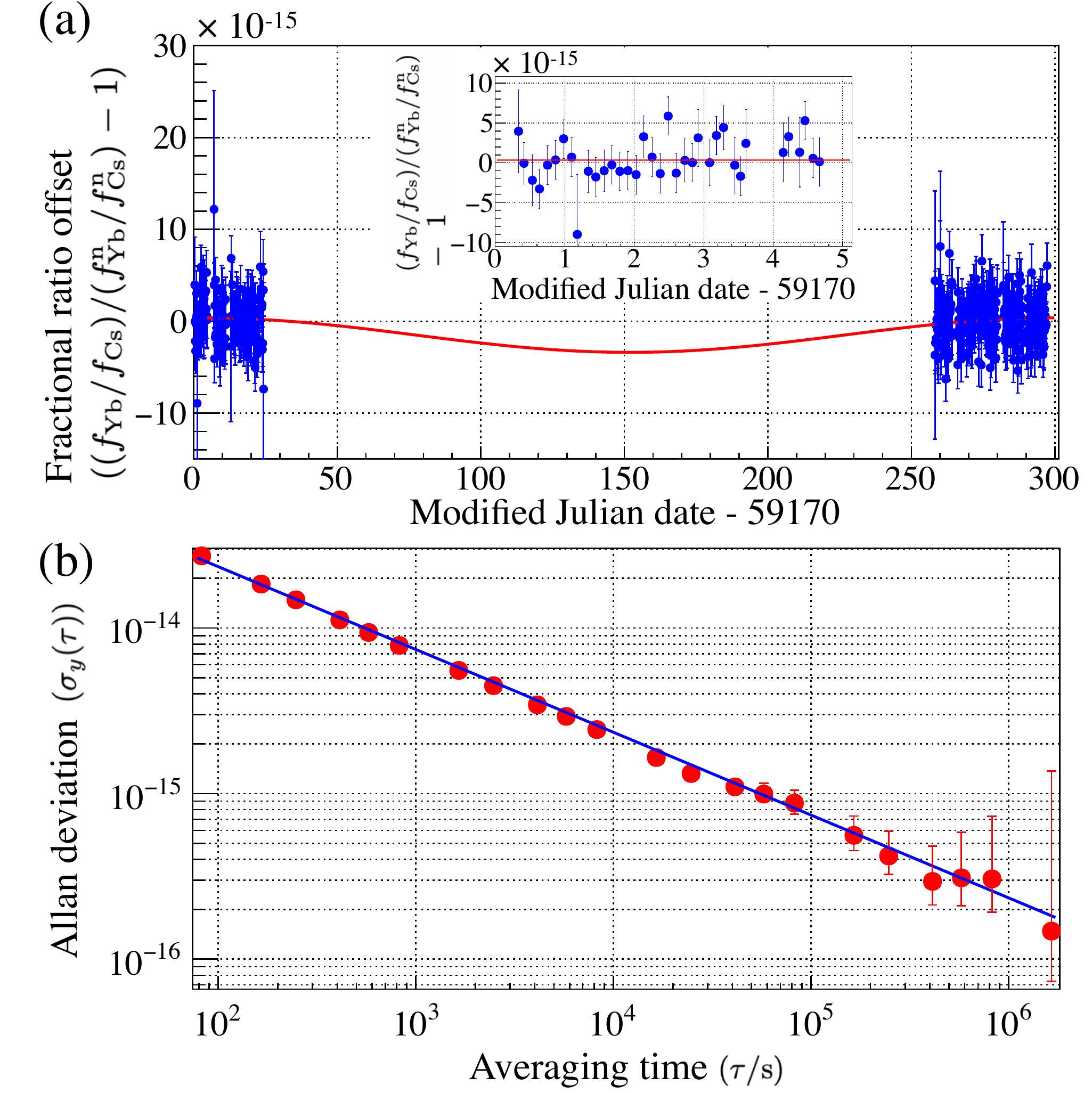}
\caption{(a) Fractional frequency ratio of Yb/Cs averaged over an interval of $1\times10^{4}$ s from Modified Julian date 59170 (17th November 2020). $(f_{\mathrm{Yb}}/f_{\mathrm{Cs}})/(f_{\mathrm{Yb}}^{\mathrm{n}}/f_{\mathrm{Cs}}^{\mathrm{n}})-1$ in the vertical axis denotes the fractional offset of the measured ratio from the ratio calculated by nominal frequencies $f_{\mathrm{Cs}}^{\mathrm{n}}=9\,192\,631\,770$ Hz and  $f_{\mathrm{Yb}}^{\mathrm{n}}=518\,295\,836\,590\,863.6$ Hz \cite{Riehle2018}. The error bar indicates the combined statistical and systematic uncertainty of the frequency measurement (see Table \ref{budget}), dominated by the statistical uncertainty. The red curve shows an example of the fit of a sinusoidal function. The inset shows an enlarged view of the data points. (b) Allan deviation of the frequency ratio calculated from concatenated data. The blue line indicates a slope of $2.3\times10^{-13}/\sqrt{(\tau /\mathrm{s})}$.}
\label{dataplot}
\end{figure}

During a search period of $T_{\mathrm{total}}=298$ days from 17th November 2020 to 11th September 2021, the measurement of the frequency ratio Yb/Cs was carried out in two campaigns with periods of (i) 25 days from 17th November 2020 and (ii) 40 days from 2nd August 2021, with uptimes of 64.4 $\%$ and 74.5 $\%$, respectively. Figure \ref{dataplot} (a) shows the data points of the fractional frequency ratio. The data were averaged over an interval of $1\times10^{4}$ s, since this work aims to search for oscillations at low frequencies. The total number of the averaged data was $N_{\mathrm{data}}=447$. Figure \ref{dataplot} (b) shows the Allan deviation of the ratio measurement, indicating the domination of the white frequency noise in the Yb/Cs data.

\begin{figure}[t]
\includegraphics[scale=0.39,angle=0]{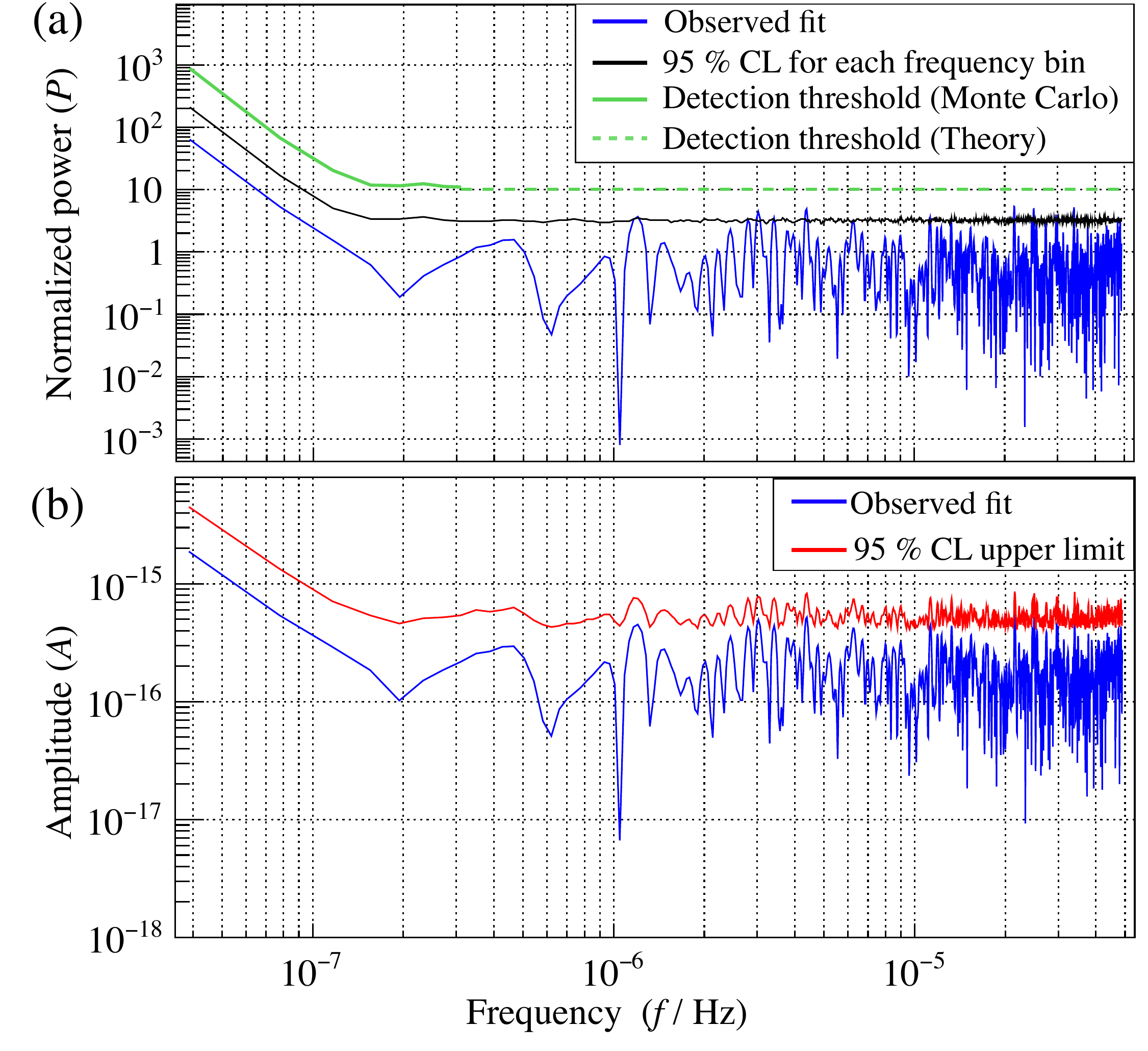}
\caption{(a) Normalized power spectrum $P=A^{2}N_{\mathrm{data}}/(4\sigma^{2})$ obtained from the fit (blue line). The black line shows the 95 $\%$ confidence level (CL) of the simulated noise spectrum for each frequency bin. The green solid and dashed lines indicate the detection thresholds (see text) calculated by a Monte Carlo simulation and a theoretical calculation, respectively. (b) Amplitude spectrum $A$ (blue line) and upper limits of $A$ at the 95 $\%$ confidence level (red line).}
\label{amplitude}
\end{figure}

To estimate the strength of harmonic oscillation signals in the measured Yb/Cs ratio, we employed an analysis method similar to those of Refs.~\cite{Tilburg2015,Hees2016,Wcislo2018,Kennedy2020}. For each oscillation frequency $f$, we carried out the chi-square fit of the ratio data in Fig.~\ref{dataplot} (a) by a function $p_{1} \cos(2\pi ft)+p_{2} \sin(2\pi ft) + p_{3}$ with free parameters $p_{1}$, $p_{2}$, and $p_{3}$, and then obtained the amplitude as $A=\sqrt{p_{1}^{2}+p_{2}^{2}}$. The analyzed frequencies were chosen from $1/T_{\mathrm{total}}\sim3.9\times10^{-8}$ Hz to $4.9 \times10^{-5}$ Hz, corresponding to the mass range from $m_{\varphi}=1.6\times10^{-22}$ eV/$c^{2}$ to $2.0\times10^{-19}$ eV/$c^{2}$. The frequency bin width was determined by $\Delta f=1/T_{\mathrm{total}}$, resulting in the total bin number of $N_{f}=1262$. We also defined the normalized power spectrum $P=A^{2}N_{\mathrm{data}}/(4\sigma^{2})$ \cite{Tilburg2015,Hees2016}, where $\sigma^{2}=(2.5\times10^{-15})^{2}$ denotes the variance of the Yb/Cs data. Figures \ref{amplitude} (a) and (b) show the obtained power $P$ and amplitude $A$ as a function of $f$, respectively (blue lines).

\begin{figure*}[t]
\includegraphics[scale=0.68,angle=0]{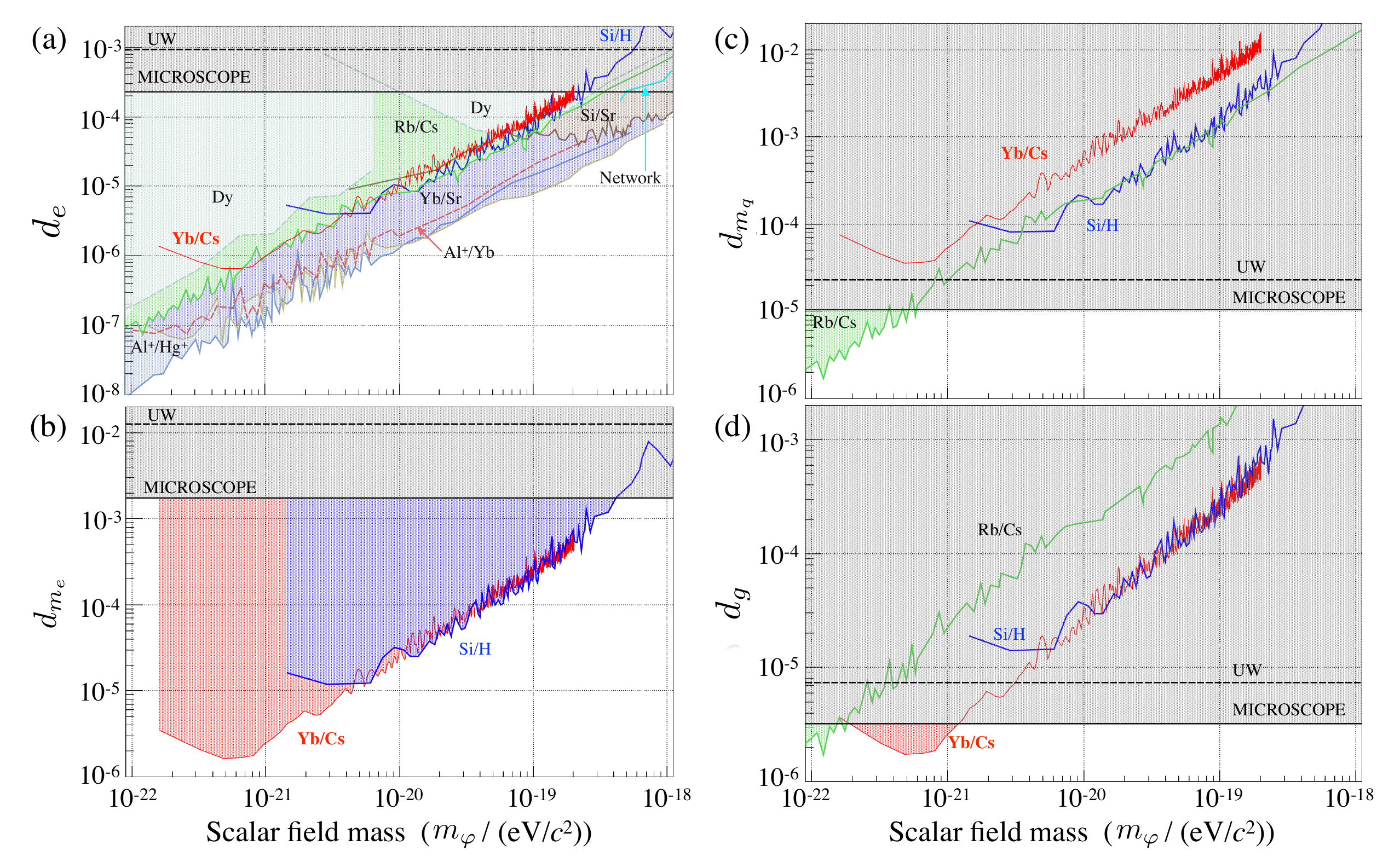}
\caption{Exclusion plots for (a) $d_{e}$, (b) $d_{m_{e}}$, (c) $d_{m_{q}}$, and (d) $d_{g}$ at the 95 $\%$ confidence level, assuming that these respective couplings dominate. The shaded areas show excluded regions set by our Yb/Cs measurement, previous atomic clock measurements of Si/Sr \cite{Kennedy2020}, Yb/Sr \cite{Beloy2021}, Al$^{+}$/Hg$^{+}$ \cite{Beloy2021}, Al$^{+}$/Yb \cite{Beloy2021}, Rb/Cs \cite{Hees2016}, Si/H \cite{Kennedy2020} ratios, an optical clock network \cite{Wcislo2018}, Dy spectroscopy \cite{Tilburg2015}, equivalence principle tests by University of Washington (UW) \cite{Schlamminger2008} and the MICROSCOPE satellite experiment \cite{Berge2018}. To take account of interferences between the scalar field oscillations \cite{Centers2021}, the limits derived from the atomic clocks and the Dy spectroscopy are rescaled by a factor of 3.0.}
\label{upperbound}
\end{figure*}

To determine whether we observed a signal exceeding a noise level, an expected noise power spectrum was calculated by a Monte Carlo simulation with a white frequency noise model \cite{supplemental}. The black line in Fig.~\ref{amplitude} (a) shows the 95 $\%$ confidence level of the noise spectrum for each frequency bin. We found several events exceeding this level, since many frequency bins ($N_{f}=1262$) were investigated. To take account of this so-called look-elsewhere effect, we defined a detection threshold \cite{Hees2016,Wcislo2018} such that the probability of finding a noise above this threshold is 5 $\%$ for a search involving $N_{f}$ bins \cite{detectionth}. The detection threshold was calculated by a Monte Carlo simulation and a theoretical calculation \cite{supplemental}, and are shown in Fig.~\ref{amplitude} (a) (green line). We observed no signals exceeding the detection threshold. Figure \ref{amplitude} (b) shows upper limits of the amplitudes at the 95 $\%$ confidence level (red line) \cite{supplemental}. These limits are model-independent results of this paper. 

Upper bounds on the coupling coefficients $d_{e}$, $d_{m_{e}}$, $d_{m_{q}}$, and $d_{g}$ were transformed from the amplitude limits using Eq.~(\ref{darkmatteramp}), assuming that these respective couplings dominate. Interferences between the scalar field oscillations arise from a velocity distribution of the scalar fields in our galaxy, resulting in a temporal variation of the field amplitude $\varphi_{0}$ \cite{Centers2021}. Since search periods by atomic clocks are typically shorter than expected coherence times of the field oscillations (e.g., $>600$ years in our target mass regions) \cite{Krauss1985}, the variation of $\varphi_{0}$ may lead to underestimation of the limits on the coupling coefficients. With an assumed probability distribution of $\varphi_{0}$ \cite{Centers2021}, the 95 $\%$ confidence limits on $d_{e}$, $d_{m_{e}}$, $d_{m_{q}}$, and $d_{g}$ were therefore rescaled by a factor of 3.0.

Figures \ref{upperbound} (a)$-$(d) show our 95 $\%$ confidence limits on the coupling coefficients as a function of $m_{\varphi}$, together with those derived from those derived from previous atomic clock measurements of Si/Sr \cite{Kennedy2020}, Yb/Sr \cite{Beloy2021}, Al$^{+}$/Hg$^{+}$ \cite{Beloy2021}, Al$^{+}$/Yb \cite{Beloy2021}, Rb/Cs \cite{Hees2016}, Si/H \cite{Kennedy2020} ratios, an optical clock network \cite{Wcislo2018}, Dy spectroscopy \cite{Tilburg2015}, and also equivalence principle tests in which differential accelerations between two macroscopic objects have been measured \cite{Schlamminger2008,Berge2018}.  The equivalence principle tests search for a Yukawa force mediated by the virtual exchange of the scalar field. Thus, the limits set by these tests are not affected by the interference effect of the oscillation. We set new limits on $d_{m_{e}}$ at the range from $m_{\varphi}=10^{-22}$ eV/$c^{2}$ to $10^{-20}$ eV/$c^{2}$. The most stringent limit was $d_{m_{e}}\lesssim 1.6\times10^{-6}$ at $m_{\varphi}\sim5\times10^{-22}$ eV/$c^{2}$, which was improved by a factor of $\sim10^{3}$. In spite of the relatively stringent constraints on $d_g$ by the equivalence principle tests, which is because the mass of a macroscopic object mostly consists of the nucleon mass \cite{Damour2010,Arvanitaki2015}, we improved the limits on $d_{g}$ at the range from $m_{\varphi}=10^{-22}$ eV/$c^{2}$ to $10^{-21}$ eV/$c^{2}$. While the constraints on the other coefficients $d_{e}$ and $d_{m_{q}}$ are not improved in this work, our limits are complementary to those of previous works. 

\begin{table}[b]
\caption{Uncertainty budget for the absolute frequency measurement of the $^{171}$Yb clock transition. For more details, see Ref.~\cite{supplemental}.}  
	\label{budget}
	\begin{center} 
\begin{tabular}{lc}
\hline
Contribution  &  Fractional uncertainty$/10^{-16}$ \\
\hline
Systematics \\
NMIJ-Yb1  & 1.4  \\
NMIJ-F2 & 4.7 \\
Gravitational & 0.1 \\
Link & 1.8  \\
\hline
Statistics & 1.2\\
\hline
Total &  5.3\\
\hline
\end{tabular}
\end{center}
\end{table}

Since the frequency of NMIJ-F2 realizes the definition of the SI second, the measured Yb/Cs ratio yields the absolute frequency of Yb. From the weighted mean value of the ratio in Fig.~\ref{dataplot} (a), we obtained the absolute frequency 
\begin{equation}
f_{\mathrm{Yb}}=518\,295\,836\,590\,863.69(28)\,\mathrm{Hz}
\end{equation}
with a fractional uncertainty of $5.3\times10^{-16}$. The uncertainty budget of this measurement is shown in Table \ref{budget}. Our measured frequency was in good agreement with previous values  \cite{McGrew2019,Pizzocaro2017,Pizzocaro2020,Kobayashi2020,Luo2020,Kim2021} at the $10^{-16}$ levels.

In conclusion, we have searched for harmonic oscillation signals from long-term comparison data between NMIJ-Yb1 and NMIJ-F2. We improved constraints on $d_{m_{e}}$ in the mass range from $m_{\varphi}=10^{-22}$ eV/$c^{2}$ to $10^{-20}$ eV/$c^{2}$ and $d_{g}$ in the range of $m_{\varphi}\lesssim10^{-21}$ eV/$c^2$, assuming that these coupling strengths dominate. Especially, limits on $d_{m_{e}}$ at $m_{\varphi}\sim5\times10^{-22}$ eV/$c^{2}$ were improved by three orders of magnitude. We also provided the absolute frequency of the Yb clock transition with an uncertainty of $5.3\times10^{-16}$. This work has demonstrated that long-term operation of an optical clock is essential for a dark matter search as well as timekeeping. While only a few other groups have demonstrated the nearly continuous operations of optical clocks for $\sim1$ month \cite{Riedel2020}, we expect that the uptimes of optical clocks will increase, contributing to improve the detection sensitivity of dark matter. 

We would like to thank H. Katori, M. Takamoto, and H. Imai for providing information on their vacuum systems of the optical lattice clock. We are indebted to F.-L. Hong for helpful discussions from the early stage of the development of NMIJ-Yb1. We are grateful to K. Sugawara for the surface roughness measurement for the systematic evaluation of NMIJ-Yb1. This work was supported by Japan Society for the Promotion of Science (JSPS) KAKENHI Grant Number 15K21669, 17H01151, 17K14367, 18K04989, 22H01241, and JST-Mirai Program Grant Number JPMJMI18A1, Japan.

$Note$ $added.-$While completing the manuscript, we found recent results of the MICROSCOPE experiment that have improved the limit on the violation of the equivalence principle by a factor of 4.6 \cite{Touboul2022}.

\section*{Supplemental Material}%

\section{Systematic uncertainty evaluation of NMIJ-Yb1}
Table \ref{systematictalbe} shows the updated nominal systematic uncertainty budget of the $^{171}$ Yb optical lattice clock NMIJ-Yb1. In this Supplemental Material, the frequency shifts and uncertainties are described by fractional values relative to the Yb atomic transition frequency of 518 THz. The total systematic uncertainty is improved to $9.8\times10^{-17}$ compared with $\sim4\times10^{-16}$ in our previous evaluations \cite{Kobayashi2018,Kobayashi2020}. 

\begin{table}[t]
\caption{Updated nominal systematic uncertainty budget of NMIJ-Yb1. BBR: blackbody radiation, AOM: acousto-optic modulator.}  
	\label{systematictalbe}
	\begin{center} 
\begin{tabular}{lcc}
\hline
Effect  & $\mathrm{Shift}/10^{-17}$ & $\mathrm{Uncertainty}/10^{-17}$ \\
\hline
Lattice light & 3.6 & 5.6 \\
BBR & $-254.0$ & 7.7 \\
Density & $-2.2$ & 1.2 \\
Second-order Zeeman & $-5.1$ & 0.3 \\
Probe light & 0.4 & 0.2 \\
Servo error & $-1.2$ & 1.2\\
AOM switching & $-$& 1 \\
Line pulling & $-$& 1 \\
DC Stark & $-$& $0.1$ \\
\hline
Total & $-258.5$ & 9.8\\
\hline
\end{tabular}
\end{center}
\end{table}

\subsection{Lattice light shift}
To evaluate the lattice light shift, we employed a light shift model \cite{Katori2015,Nemitz2019} that includes the contribution from the electric-dipole ($E1$) polarizability, the multipolar ($M1$ and $E2$) polarizabilities, and hyperpolarizability. When the lattice laser frequency $f_{\mathrm{L}}$ is detuned from the $E1$ magic frequency $f_{E1}$ by $\Delta f$, the light shift $\Delta f_{\mathrm{LS}}$ is given by
\begin{eqnarray}
\Delta f_{\mathrm{LS}}&=&(\alpha\Delta f - \alpha_{\mathrm{qm}})\Big(\Braket{n}+\frac{1}{2}\Big)\Big(U_{\mathrm{e}}/E_\mathrm{r}\Big)^{1/2}\nonumber\\
&&+\beta(2\Braket{n}+1)\Big(U_{\mathrm{e}}/E_{\mathrm{r}}\Big)^{3/2}\nonumber\\
&&-\Big\{\alpha\Delta f + \frac{3}{4}\beta\Big(2\Braket{n}^{2}+2\Braket{n}+1\Big)\Big\}U_{\mathrm{e}}/E_\mathrm{r}\nonumber\\
&&-\beta\Big(U_{\mathrm{e}}/E_{\mathrm{r}}\Big)^2,
\label{lightshifteq}
\end{eqnarray}
where $U_{\mathrm{e}}$ is the effective trap depth which is less than the maximum trap depth $U_{0}$ due to the radial motion of the trapped atoms, $E_{\mathrm{r}}$ the lattice photon recoil energy used as a conventional unit of the trap depth, and $\Braket{n}$ the average vibrational quantum number. $\alpha$, $\alpha_{\mathrm{qm}}$, and $\beta$ denote the $E1$ polarizability, the multipolar polarizability, and hyperpolarizability coefficients, respectively. These atomic coefficients were taken from values reported in Ref.~\cite{Nemitz2019}. The effective depth was calculated by $U^{m}_{\mathrm{e}}\sim\{(\xi+\delta_{m})U_{0}\}^{m}$, where $m$ denotes the exponent (1/2, 3/2, 1, or 2), $\xi$ a reduction factor of the trap depth, and $\delta_{m}$ a small correction for $m\neq1$. $\delta_{1/2}$ and $\delta_{3/2}$ can be eliminated by the relationships $\delta_{1/2}\sim-\frac{1}{2}\delta_{2}$ and $\delta_{3/2}\sim\frac{1}{2}\delta_{2}$ \cite{Nemitz2019}. The parameters $U_{0}$, $\Braket{n}$, $\xi$, and $\delta_{2}$ were estimated by sideband spectroscopy of the clock transition \cite{Blatt2009}.

To find $f_{E1}$, we carried out an interleaved measurement in which the light shift at our operating lattice frequency $f_{\mathrm{L}}=394\,798\,263.4$ MHz was measured by the alternating stabilization of the clock laser to the atomic transition with two different trap depths (deep and shallow). Table \ref{sidebandtable} summarizes the trap parameters in this measurement. The lattice light was generated by a titanium sapphire laser (M Squared, SolsTiS-4000-SRX-R). While its background spectrum from amplified spontaneous emission is expected to be very low \cite{Fasano2021}, spectral filtering of the laser output was carried out by a volume Bragg grating with a bandwidth of 200 GHz. The atoms were always loaded to the lattice at a full lattice power, and afterwards the power was reduced to get a final trap depth. When probing the atoms in deep lattices, the trap depth was first reduced to $U_{0}\sim200E_{\mathrm{r}}$ and then increased to the final values $U_{0}\sim500E_{\mathrm{r}}$, blowing out high energy atoms \cite{Falke2014}. 

\begin{figure*}[t]
\includegraphics[scale=0.5]{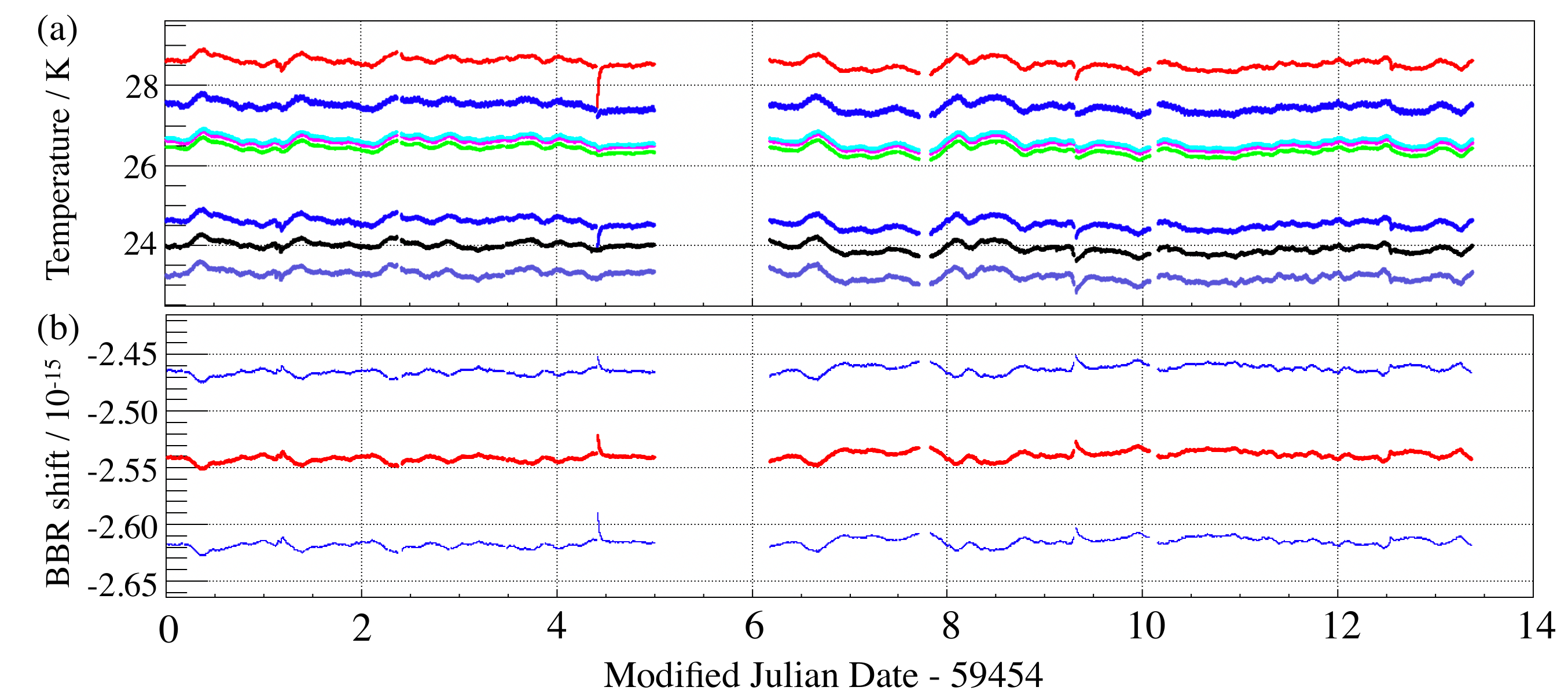}
\caption{(a) Example of temperatures of eight points in the main vacuum chamber measured during a clock operation period of 14 days from Modified Julian date 59454 (28th August 2021). Relatively large changes of the temperatures on the fourth day were caused by stopping a current applied to the magneto optical trap coil for maintenance of the clock. (b) BBR shift (red line) and its total uncertainty (blue lines). }
\label{tempmeasure}
\end{figure*}

\begin{table}[t]
\caption{Trap parameters in the interleaved measurement. The uncertainties of the parameters were estimated from the reproducibilities in several measurements.}   
	\label{sidebandtable}
	\begin{center} 
\begin{tabular}{ccccc}
\hline
 & $U_{0}/E_\mathrm{r}$ & $\Braket{n}$ & $\xi$ & $\delta_{2}$ \\
\hline
 Deep & 498(14) & 1.2(2) & 0.86(2) & 0.006(2)\\
 Shallow & 158(4) & 1.7(3) & 0.66(5) & 0.04(2)\\
 \hline
\end{tabular}
\end{center}
\end{table}

In the interleaved measurement, the shift value was obtained as $23.7(13.2)\times10^{-17}$ after $\sim6\times10^{4}$ s of averaging. Since the modulation of the trap depths also changes the density shift \cite{Nicholson2015}, the shift value was corrected according to our evaluation results of the density shift (see Section \ref{densityshiftsection}). By solving Eq.~(\ref{lightshifteq}), we obtained $f_{E1}=394\,798\,255.0(9.0)$ MHz, which agreed with the most accurate previous value \cite{Nemitz2019}. With a typical operating trap depth of $U_{0}\sim200E_{\mathrm{r}}$, the uncertainty of the lattice light shift is $\sim6\times10^{-17}$.

\subsection{Blackbody radiation shift}
\label{bbrshiftevaluationsection}
The blackbody radiation (BBR) shift $\Delta f_{\mathrm{BBR}}$ caused by an environment with a temperature of $T$ is given by
\begin{equation}
\Delta f_{\mathrm{BBR}} = \alpha_{\mathrm{stat}}\Big(\frac{T}{300\,\mathrm{K}}\Big)^{4}\Big[1+\eta_{1}\Big(\frac{T}{300\,\mathrm{K}}\Big)^{2}+\eta_{2}\Big(\frac{T}{300\,\mathrm{K}}\Big)^{4}\Big],
\label{bbrequation}
\end{equation}
where $\alpha_{\mathrm{stat}}$ denotes the coefficient resulting from the static polarizability and the Stefan-Boltzmann law, and the terms including $\eta_{1}$ and $\eta_{2}$ provide small dynamic corrections. These coefficients  were taken from values reported in Refs.~\cite{Sherman2012,Beloy2014}.
Given Eq. (\ref{bbrequation}), one needs to find the effective temperature $T_{\mathrm{eff}}$ seen by the atoms. This is determined from surrounding surfaces such that
\begin{equation}
T_{\mathrm{eff}}^{4} = \sum_{i}\Big( \frac{\Omega_{i}^{\mathrm{eff}}}{4\pi} \Big)T_{i}^{4},
\label{effectivesum}
\end{equation}
where $T_{i}$ and $\Omega_{i}^{\mathrm{eff}}$ denote the temperature and the effective solid angle of a surface $i$ \cite{Beloy2014}. The effective solid angle generally differs from the geometric solid angle, since the BBR photons reach the atoms after being multiply scattered on chamber surfaces that are characterized by their emissivities and roughnesses. For evaluation of $T_{\mathrm{eff}}$, we divided surfaces into two parts: (i) a main vacuum chamber for trapping atoms with a large solid angle of $\Omega_{\mathrm{main}}^{\mathrm{eff}} / (4\pi)\sim0.98$, and (ii) other components including hot sources such as an atomic oven and a heated window \cite{Kobayashi2020}.

The dominant contribution from the main chamber was estimated by measuring the temperatures of the chamber surfaces including an in-vacuum bobbin for the magneto optical trap (MOT) coil with eight platinum-resistance thermometers. Figure \ref{tempmeasure} (a) shows an example of the measured temperatures. The hottest point was located at the coil bobbin, while the coldest point was at the chamber farthest from the hot sources. As a probability distribution of the temperature, we employed a rectangular distribution between the maximum and minimum temperatures ($T_{\mathrm{max}}$ and $T_{\mathrm{min}}$). This gives an average temperature of $(T_{\mathrm{max}}+T_{\mathrm{min}})/2$ with an uncertainty of $(T_{\mathrm{max}}-T_{\mathrm{min}})/\sqrt{12}$ \cite{GUM}.

\begin{figure}[t]
\includegraphics[scale=0.39]{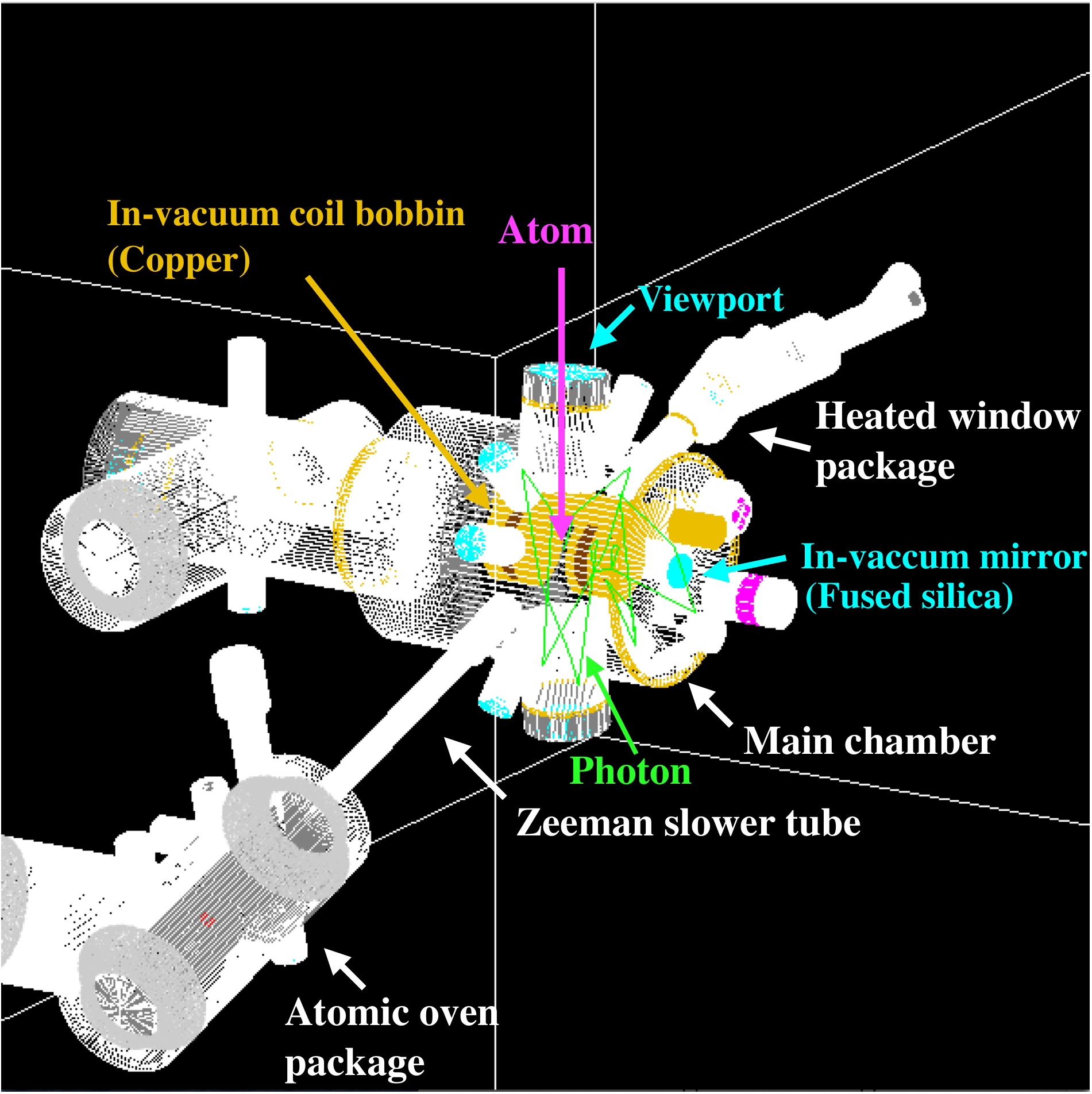}
\caption{Chamber geometry for the Monte Carlo simulation of the BBR photon trajectory (green line) based on the Geant4 simulation toolkit. The initial design of this chamber system was made by RIKEN \cite{Takamoto2020,Ohmae2021}.}
\label{geant4geometry}
\end{figure}

The contributions from the other components were evaluated by a Monte Carlo tracking of the BBR photons based on the Geant4 simulation toolkit \cite{geant4}. Figure \ref{geant4geometry} shows a schematic diagram of the chamber geometry. The photons are generated from the position of the atoms and emitted in random directions \cite{Bothwell2019}. Each photon is reflected or absorbed in each surface with a probability determined by its emissivity and roughness. We took the emissivities from literatures and set conservative uncertainties as summarized in Table \ref{emissivity}. The surface roughness is characterized by a microfacet model in which the surface is composed of many microscopic planes called microfacets. We generated $10^{7}$ photons and recorded the surface at which each photon was absorbed. The ratio of the number of photons absorbed in a surface over the generated photon number of $10^{7}$ yields the effective solid angle of the surface.  

\begin{table}[t]
\caption{Emissivities and their uncertainties employed in the Monte Carlo simulation. The emissivity of stainless-steel was chosen from values for highly polished materials, taking account of measured arithmetic average roughness $R_{\mathrm{a}}$ between 0.1 $\mu$m and $0.2$ $\mu$m (see text).}   
	\label{emissivity}
	\begin{center} 
\begin{tabular}{lcc}
\hline
Material & Emissivity & Ref.\\
\hline
Stainless-steel & $0.1^{+0.2}_{-0.05}$ & \cite{Guideline,Wieting1979,Dolezal2015,Woods2014}\\
Copper & $0.05^{+0.15}_{-0.05}$ & \cite{Guideline,Dolezal2015,Fernandez2014}\\
Fused silica & $0.8^{+0.2}_{-0.2}$ &  \cite{Dolezal2015}\\
\hline
\end{tabular}
\end{center}
\end{table}

In Geant4, the angular distribution of the microfacets relative to a macroscopic mean surface is given by a Gaussian distribution with its standard deviation $\sigma_{\mathrm{a}}$ determined by users. In this simulation, $\sigma_{\mathrm{a}}$ was determined from our measurements. The chamber surface is mostly made of stainless-steel and finished with buff and electro polishings. We prepared several stainless-steel samples and measured the roughnesses of them by a stylus-type surface roughness instrument. The arithmetic average roughness $R_{\mathrm{a}}$ was found to be between $0.1$ $\mu$m and $0.2$ $\mu$m, which was obtained after applying a Gaussian bandpass filter to surface profiles with short and long cutoff wavelengths of $2.5$ $\mu$m and 0.8 mm, respectively. In the derivation of the angular distribution, we chose a short cutoff wavelength equal to the wavelength of the BBR photon, assuming that the photon does not see a structure smaller than its wavelength. Taking account of a broad spectrum of the BBR photons, we derived the standard deviations $\sigma_{\mathrm{a}} ({\lambda_{\mathrm{c}}})$ with various short cutoff wavelengths $\lambda_{\mathrm{c}}$, and calculated a weighted mean of  $\sigma_{\mathrm{a}} ({\lambda_{\mathrm{c}}})$ with its weight determined from the BBR spectrum. As a result, $\sigma_{\mathrm{a}}$ was obtained as $0.013^{+0.022}_{-0.013}$ rad. The uncertainty was estimated from variations in different measurement positions and samples.

\begin{table}[t]
\caption{Effective solid angles of the surfaces calculated by the Monte Carlo simulation and their temperatures.}   
\label{effectivesolidtable}
	\begin{center} 
\begin{tabular}{lcc}
\hline
Surface $i$  & $\Omega_{i}^{\mathrm{eff}}/(4\pi)$ & $T_{i}/\mathrm{K}$ \\
\hline
Heated window & $0.00371$ & 485(20)\\
Copper mount of the heated window & $0.00023$ &499(10)\\
Tube between the heated window and  & $ 0.00411$ & 300.3(1.1)\\
the main chamber & & \\
Atomic oven & $0.00074$ & 644(10)\\
Radiation shield for the atomic oven  & 0.00017 & 479(101)\\
Zeeman slower tube & $0.00609$ & 303.4(7.3)\\
\hline
\end{tabular}
\end{center}
\end{table}

\begin{table}[t]
\caption{Uncertainty budget for the effective temperature $T_{\mathrm{eff}}$ based on the temperature measurement during the 14 days period in Fig.~\ref{tempmeasure}.}   
\label{tempbudget}
	\begin{center} 
\begin{tabular}{lc}
\hline
Contribution  & $\mathrm{Uncertainty}/\mathrm{K}$\\
\hline
Temperature measurement &\\
Main Chamber & 1.705 \\
Heated window & 0.307 \\
Copper mount of the heated window & 0.011 \\
Tube between the heated window and  & 0.005\\
the main chamber & \\
Atomic oven & 0.072\\
Radiation shield for the atomic oven & 0.070 \\
Zeeman slower tube & 0.045 \\
\hline
Emissivity & \\
Stainless-steel & 1.176\\
Copper & 0.083 \\
Fused silica & 0.237 \\
\hline
Surface roughness & 0.612 \\
\hline
Atom position & 0.569 \\ 
\hline
Other contributions & 0.248\\
\hline
Total & 2.284\\
\hline
\end{tabular}
\end{center}
\end{table}

\begin{figure}[t]
\includegraphics[scale=0.35]{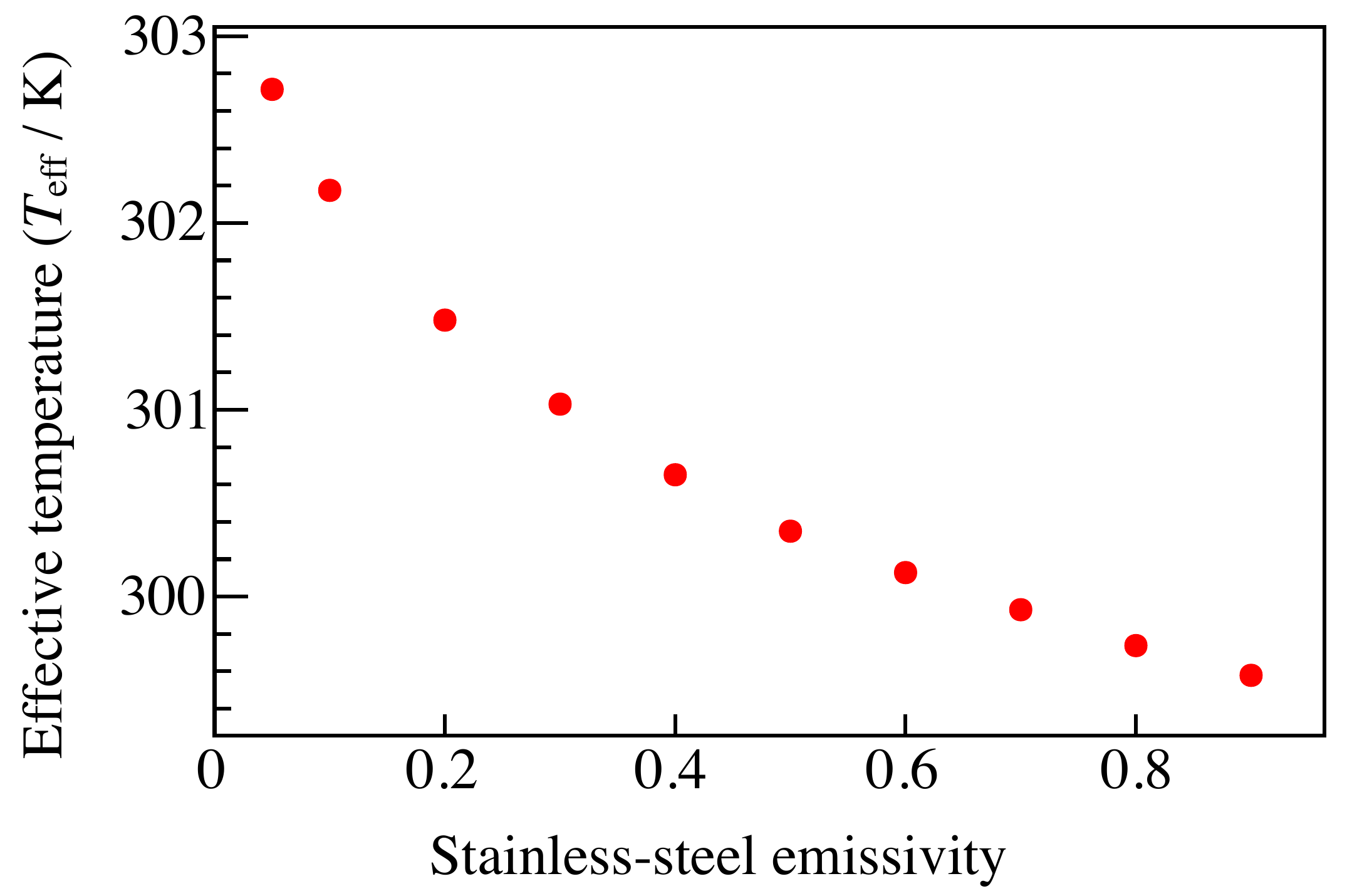}
\caption{Effective temperature as a function of the stainless-steel emissivity.}
\label{stainless}
\end{figure}

Table \ref{effectivesolidtable} lists some examples of the effective solid angles obtained by the Monte Carlo simulation. The effective temperature $T_{\mathrm{eff}}$ is obtained by adding these contributions to the contribution from the main chamber according to Eq. (\ref{effectivesum}). We found that $T_{\mathrm{eff}}$ is shifted by $\sim3$ K from the effective temperature determined only from the main chamber ($\sim299.2$ K). To evaluate the uncertainty of $T_{\mathrm{eff}}$ arising from the uncertainties of the emissivity and the surface roughness, the dependences of $T_{\mathrm{eff}}$ on these surface parameters were investigated. Figure \ref{stainless} shows an example of $T_{\mathrm{eff}}$ as a function of the stainless-steel emissivity.

Table \ref{tempbudget} shows the uncertainty budget for $T_{\mathrm{eff}}$. The total uncertainty is $2.3$ K, corresponding to the BBR frequency shift uncertainty of $7.7\times10^{-17}$. The uncertainty is mostly due to the temperature inhomogeneity in the main chamber and the uncertainty of the stainless-steel emissivity. Other contributions in the last line of Table \ref{tempbudget} arise from components with small solid angles such as a vacuum pump and tubes that cover the heated window and the atomic oven. Since the temperatures of the chamber slowly drift, the BBR shift and its total uncertainty slightly change as a function of time (see Fig.~\ref{tempmeasure} (b)). The uncertainty is conservatively taken from the maximum value in the measurement period. 

In the first Yb/Cs measurement campaign from 17th November 2020 (see the main text), the BBR shift uncertainty was relatively large ($1.3\times10^{-16}$). This was mostly because the temperature inhomogeneity in the main chamber was large ($T_{\mathrm{max}}-T_{\mathrm{min}}\sim10$ K) due to imperfect heat dissipation of the MOT coil bobbin. 

\subsection{Density shift}
\label{densityshiftsection}
\begin{figure}[t]
\includegraphics[scale=0.3]{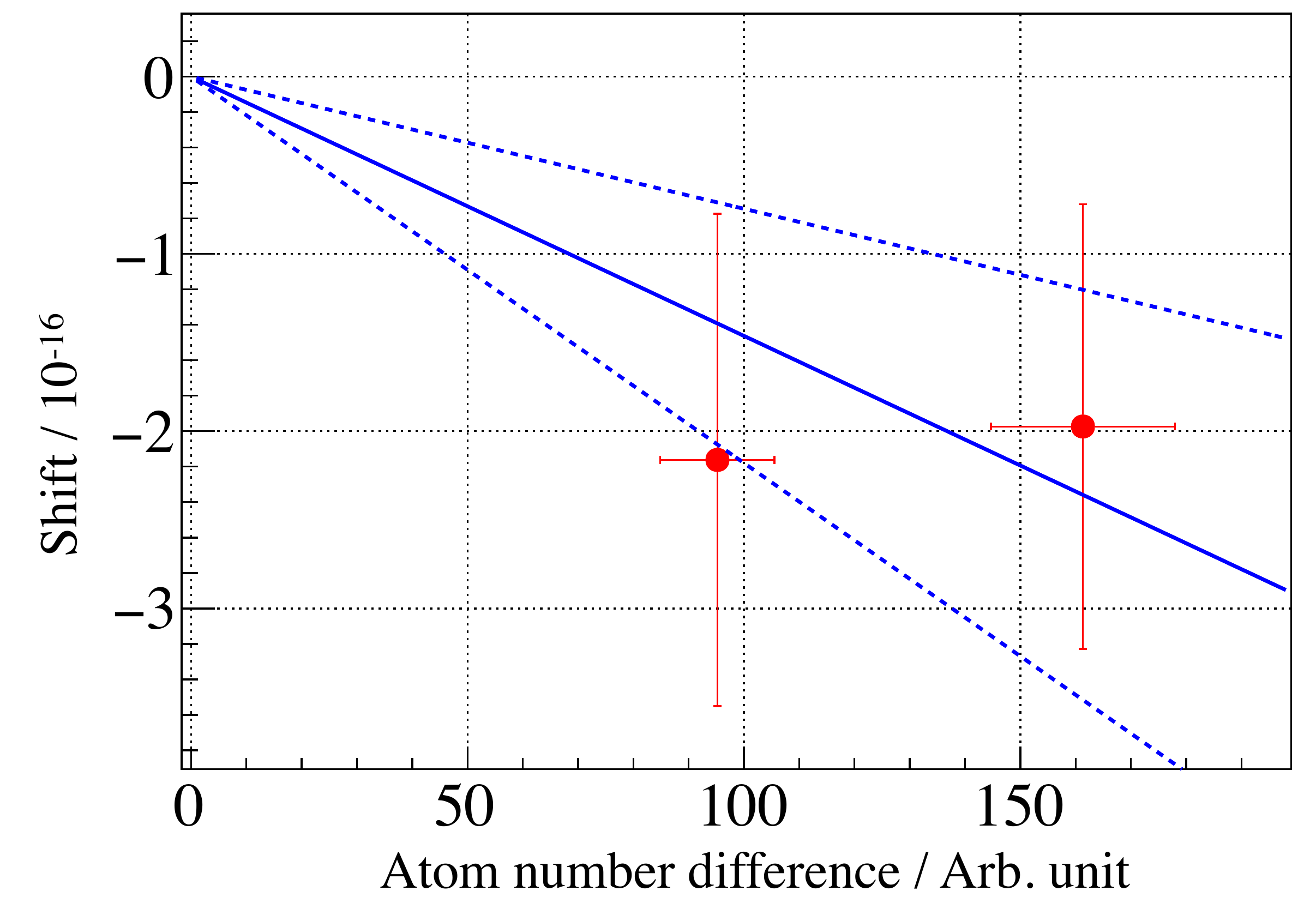}
\caption{Density shift as a function of the atom number difference. The solid (dashed) blue line indicates the linear fit (its uncertainty).}
\label{densityshift}
\end{figure}

The density shift was reevaluated with an interleaved measurement in which the number of trapped atoms $N_{\mathrm{atom}}$ was varied. Figure \ref{densityshift} shows the measured shift as a function of the atom number difference. The coefficient of the density shift was obtained by fitting the measured data with a linear function. In this measurement, the trap depth and atomic temperature were $U_{0}=646(18)E_{\mathrm{r}}$ and $T_{\mathrm{a}}=17(3)$ $\mu$K, respectively. When calculating the density shift at our operating density, we used a scaling of the density shift with $U_{0}^{3/2}/\sqrt{T_{\mathrm{a}}}$ \cite{Nicholson2015,Nicholsonthesis}. For a typical operating condition of $N_{\mathrm{atom}}\sim30$ (arb.~unit), $U_{0}\sim200E_{\mathrm{r}}$, and $T_{\mathrm{a}}\sim9$ $\mu$K, the uncertainty of the density is $\sim1\times10^{-17}$.

\subsection{Other shifts}
The evaluations of the second-order Zeeman shift, probe light shift, servo error, and AOM (acousto-optic modulator) switching are same as those in Refs. \cite{Kobayashi2018,Kobayashi2020}.

The uncertainty of the line pulling arising from possible imperfect spin-polarization was conservatively estimated as $1\times10^{-17}$.

The stray electric fields on dielectric materials such as an in-vacuum mirror and viewports (see Fig.~\ref{geant4geometry}) cause the DC Stark shift $\Delta f_{\mathrm{DC}}$. We carried out a finite element analysis to estimate the electric field at the positions of the atoms. The electric field from the in-vacuum mirror is attenuated by a factor of $10^{5}$, thanks to the shielding of the coil bobbins. Assuming that the mirror initially traps maximum possible charges that generate the breakdown electric field of air ($\sim3$ MV/m), the absolute value of the DC Stark shift $|\Delta f_{\mathrm{DC}}|$ was estimated to be less than $1\times10^{-17}$ using the reported static polarizability \cite{Sherman2012}. The charge trapped on the mirror is expected to decay through its fused silica substrate \cite{Lodewyck2012}. With our estimated decay time constant of $\sim800$ days, $|\Delta f_{\mathrm{DC}}|$ was calculated to be $<1\times10^{-18}$. We also measured the DC electric potentials on the viewports with an electrostatic sensor (Keyence SK-050), and found them to be less than $\sim10$ V (measurement uncertainty of 10 V). With this result, the finite element analysis indicates that the contributions from the viewports are negligibly small.

\section{Statistical analysis of the power spectrum and derivation of the amplitude limit}
The 95 $\%$ confidence level of the noise spectrum for each frequency bin (see Fig.~3 (a) of the main text) was calculated with a Monte Carlo simulation. Since the Yb/Cs ratio is dominated by the white frequency noise, we generated simulation data that only include the white frequency noise with its standard deviation determined from the experimental data. The simulated data were gapped according to the dead time of the experimental data, and fitted with the sinusoidal function to obtain the normalized power $P$. We repeated this procedure $10^{3}$ times and obtained a probability distribution function of $P$ for each frequency bin. The 95 $\%$ confidence level $P_{0.95}$ was obtained by requiring that the cumulative probability distribution function $\mathrm{CDF}(P\le P_{0.95})$ is equal to 0.95. 
For high frequencies $f\gtrsim1/T_{\mathrm{data}}$, where $T_{\mathrm{data}}\sim4.0\times10^{6}$ s is the total data duration, the statistical properties of the normalized power spectrum can be approximately described by a theoretical probability distribution for white frequency noise, i.e., $\mathrm{CDF}(P\le P_{0.95})=1-\exp(-P_{0.95})$ which is independent of $f$ \cite{Scargle1982}. We checked that the simulated 95 $\%$ confidence level is consistent with a theoretical estimate of $P_{0.95}=-\ln(1-0.95)\sim3.0$. 

In the above definition of the 95 $\%$ confidence level, it is likely that one finds 1 event exceeding this level per 20 frequency bins. Several such events were observed in Fig. 3 (a) of the main text, since we looked for many frequency bins ($N_{f}=1262$). To take account of this look-elsewhere effect, we defined the detection threshold $P_{\mathrm{th}}$ such that the probability of finding a noise above $P_{\mathrm{th}}$ is 5 $\%$ for a search involving $N_{f}$ bins. This requires $\mathrm{CDF}(P\le P_{\mathrm{th}})=0.95^{1/N_{f}}$ for each frequency bin. The detection threshold was estimated by a Monte Carlo simulation for $f\lesssim1/T_{\mathrm{data}}$ and a theoretical calculation $P_{\mathrm{th}}=-\ln(1-0.95^{1/N_{f}})\sim10$ for $f\gtrsim1/T_{\mathrm{data}}$. 

Upper limits on the amplitudes at the 95 $\%$ confidence level were derived assuming that the observed amplitudes are true signals. We generated another Monte Carlo simulation dataset including the white frequency noise and signals. The 95 $\%$ confidence upper limits were calculated from obtained probability distributions.

\section{Uncertainty of the absolute frequency measurement}
The systematic uncertainty of NMIJ-Yb1 was evaluated as $1.4\times10^{-16}$ in the first campaign from 17th November 2020 and $9.8\times10^{-17}$ in the second campaign from 2nd August 2021. This difference arose from the difference in the uncertainty of the BBR shift as described in Section \ref{bbrshiftevaluationsection}. We conservatively took the larger uncertainty of $1.4\times10^{-16}$ in the whole measurement. The systematic uncertainty of NMIJ-F2 was evaluated as $4.7\times10^{-16}$ \cite{Takamizawa2021}. The uncertainty of the gravitational redshift due to the height difference between the two clocks ($\sim7$ m) was conservatively estimated to be $1\times10^{-17}$. 

We evaluated the link uncertainty as follows. An uncertainty arises from phase variations of a 10 MHz signal from UTC(NMIJ) that occurred during its transmission through a coaxial cable between UTC(NMIJ) and NMIJ-Yb1. This uncertainty was conservatively evaluated as $1.0\times10^{-16}$. The uncertainty of the measurement of the AOG frequency was estimated to be $2.7\times10^{-17}$ from measurement noise of the time-interval counter. We also measured the frequency difference between UTC(NMIJ) and the H maser after sending their signals to the location of NMIJ-F2 via cables, and compared it with the AOG frequency measured near the H maser. From this comparison, we evaluated the link uncertainty between UTC(NMIJ) and NMIJ-F2 as $1.5\times10^{-16}$. The time stamps of the comparison data between (i) NMIJ-Yb1 and the H maser and (ii) NMIJ-F2 and the H maser were not completely synchronized due to their different cycle times. The uncertainty due to this effect was estimated to be $1.7\times10^{-17}$ by a Monte Carlo simulation using a noise model of the H maser \cite{Kobayashi2020}. In our previous experiments \cite{Kobayashi2020}, the link uncertainty was $2.2\times10^{-16}$ which mostly arose from frequency multiplication of the 10 MHz signal. In this work, this uncertainty was improved to $<2\times10^{-17}$ by carefully stabilizing the temperature of a frequency multiplier. In total, the link uncertainty was evaluated to be $1.8\times10^{-16}$.


\begin{thebibliography}{1}
\bibitem{detectionth} In other words, if the search involving $N_{f}$ bins is repeated 20 times, one finds 1 event exceeding the detection threshold.
\end{thebibliography}
\end{document}